\documentclass[%
twocolumn,
 amsmath,amssymb,
 aps,
 pre,
]{revtex4}
\usepackage[usenames,dvipsnames]{xcolor}
\usepackage{amsmath}  
\usepackage{lipsum}
\usepackage{dsfont}
\usepackage{braket}
\usepackage{graphicx}
\usepackage{dcolumn}
\usepackage{bm}
\usepackage[colorlinks=true,citecolor=blue,linkcolor=red]{hyperref}
\usepackage[mathlines]{lineno}

\DeclareMathOperator{\tr}{Tr}
\begin{document}

\title{Thermodynamically consistent master equation based on subsystem eigenstates\footnote{Phys. Rev. E 107, 014108 (2023)}}

\author{Si-Ying Wang$^{1}$}
\author{Qinghong Yang$^{2}$}\email[Corresponding Author: ]{ yqh19@mails.tsinghua.edu.cn}
\author{Fu-Lin Zhang$^{1}$}\email[Corresponding Author: ]{ flzhang@tju.edu.cn}

\affiliation{$^{1}$Department of Physics, School of Science, Tianjin University, Tianjin 300072, China\\
$^{2}$State Key Laboratory of Low Dimensional Quantum Physics, Department of Physics, Tsinghua University, Beijing, 100084, China}


\date{\today}

\begin{abstract}
Master equations under appropriate assumptions are efficient tools for the study of open quantum systems. For many-body systems,  subsystems of which locally couple to thermal baths and weakly interact with each other, the local approach provides a more convenient description than the global approach. However, these local master equations are believed to generate inconsistencies with the laws of thermodynamics when intersubsystem interactions exist. Here we develop an alternative local master equation  by virtue of similar approximations used in deriving the traditional Gorini-Kossakowski-Lindblad-Sudarshan master equation. In particular, we stick to using  eigenstates of each subsystem to construct quantum jump operators, and the secular approximation is also employed to modify the intersubsystem interactions. Our results show that violations of thermodynamic laws will be avoided after correcting intersubsystem interactions. Finally, We  study a two-qubit heat transfer model and this further shows the validity of our modified master equation.
\end{abstract}

\maketitle

\section{Introduction}
The theoretical description of open quantum systems is a subject of both fundamental and practical importance. On the one hand, quantum systems can not be completely isolated from the environment; on the other hand, external apparatuses are needed to manipulate and control the quantum system of interest. There have already been various well-established treatments of open quantum systems~\cite{open1,open2,open3,open4}, and they have been applied to many fields, such as quantum optics \cite{open3,APPLY1}, quantum thermodynamics \cite{apply2,apply4,apply6,apply5}, quantum chemistry \cite{apply7,che}, and quantum information \cite{nielsen,apply3}.

For Markovian cases where memory effects of the environment can be neglected \cite{markov1,markov2,markov3,markov4,markov5}, the Gorini- Kossakowski-Lindblad-Sudarshan (GKLS) master equations are widely used to describe the dynamics of quantum systems~\cite{GKLS1,GKLS2,GKLS3,GKLS4,GKLS5,GKLS6,GKLS7}. For quantum systems which are composed of several interacting subsystems, two main approaches~\cite{glocal0,glocal1,glocal2,glocal3,glocal4,glocal5,naseem,glocal6,glocal7} --- the global approach and the local approach --- are developed to describe the time evolution.
The global approach is based on eigenstates of the whole Hamiltonian which is composed of the Hamiltonian of each subsystem and the interaction between subsystems.  Although it is thermodynamically consistent, those eigenstates and eigenvalues are difficult to get, especially for large-size systems. In contrast, only the eigenstates of each individual subsystem are needed for the local approach, which holds in the weak intersubsystem coupling regime~\cite{glocal0}. However, the local approach was criticized for violating the thermodynamic laws. In Refs. \cite{violate2002,violation1}, the authors show that the local approach will result in heat currents spontaneously flowing from the cold bath to the hot bath, and thus violates the second law of thermodynamics. The origin of this violation is found to result from the error of approximations made in previous local treatments~\cite{violation2}. Resorting to artificial microscopic collisional models~\cite{heat4,heat4.1,heat4.2}, one finds that after considering the extra work required to maintain collisions, the violation of the second law can be avoided. Sticking to physical models, rescaled Hamiltonians are used to derive thermodynamically consistent master equations~\cite{heat5,heat3,heat2021}, which in fact are still global master equations. In addition, Ref. \cite{heat6PRR} shows that by redefining the heat current, the violation can also be avoided.

Here, we introduce an alternative GKLS local master equation with thermodynamical consistency. In particular, we truly stick to the local picture, that is, we use eigenstates of each individual subsystem to construct the quantum jump operators and apply a local definition of the heat current. Moreover, similar approximations as in deriving the familiar GKLS master equation are used. Therefore, our equation shares the advantage of a local approach and has a similar application scope with the familiar GKLS master equation. The main idea of our work is that we apply the secular approximation ~\cite{open3,open4} to eliminate some components of the intersubsystem interaction. Those redundant components do not contribute to the time evolution of the open quantum system in the weak coupling limit, and the remaining part of the intersubsystem interaction plays the role of producing transitions in the degenerate eigensubspace of the Hamiltonian $H_s=\sum_{i=1}^nH_i$, which is the Hamiltonian of the system without intersubsystem interactions [see Eq.~\eqref{eq:TH}]. In this sense, it is natural to define the local heat current directly using the non-interacting Hamiltonian $H_s$ [see Eq.~\eqref{Q}]. Based on this definition of the local heat current, we generally show that our modified local master equation satisfies the first and second laws of thermodynamics at the same time, and then solves the problem raised in Refs. \cite{violation1,violate2002}.


\section{The modified local master equation}
We begin by considering the case that a system $\mathbb{S}$ is composed of $n$ subsystems $s_{1},s_2,\cdots,s_{n}$, which weakly interact with each other, and each of which is weakly coupled to a bath $b_{i}\,(i=1,2,\cdots,n)$. Note that this is a general case for which the local GKLS master equation is suitable. In such a scenario, the total time-independent Hamiltonian can be written as:
\begin{equation}\label{eq:TH}
	H=\sum_{i=1}^{n}(H_{i}+H_{b_i}+\alpha V_{ib_i})+\alpha H_{I},	
\end{equation}
where $H_{i}$ is the Hamiltonian of the subsystem $s_{i}$, $H_{b_i}$ is
the Hamiltonian of the bath $b_{i}$, $\alpha V_{ib_i}$ is the interaction between $s_{i}$ and $b_{i}$, and the intersubsystem interaction is denoted as $\alpha H_{I}$, which may be composed of two-subsystem interactions, three-subsystem interactions, etc. Here, $\alpha$ is a small dimensionless parameter, which accounts for the weak coupling strength, and we assume that the intersubsystem interaction and the system-bath coupling are of the same order of magnitude. However, we emphasize that our treatment is not restrictive to this assumption, and the case that strengths of $V_{ib_i}$ and $H_I$ are different, is considered in Appendix~\ref{sec:details}.

The time evolution of the whole system (including baths) is described by the von Neumann equation, which, in the Schr\"odinger picture, can be written as
\begin{align}\label{eq:vNE}
	\partial_{t}\rho(t)=-i[H,\rho(t)].
\end{align}
It is easy to derive the quantum Markovian master equation in the interaction picture, thus we transform Eq.~\eqref{eq:vNE} through the unitary transformation generated by the non-interacting Hamiltonian $H_0\equiv\sum_{i=1}^n(H_{i}+H_{b_i})$. After this transformation, Eq. \eqref{eq:vNE} becomes
\begin{equation}
	\partial_{t}{\tilde{\rho}(t)}=-i\alpha\left[\sum\limits_{i=1}^n\tilde{V}_{ib_i}
	(t)+\tilde{H}_{I}(t),\tilde{\rho}(t)\right], \label{interaction ev}
\end{equation}
where operators in the interaction picture are defined as $\tilde{O}(t)=e^{iH_0t}Oe^{-iH_0t}$ with $O$ being the operator in the Schr\"odinger picture.

Following the standard routine, we integrate Eq. \eqref{interaction ev} once, and get
\begin{equation}
	\tilde{\rho}(t)=\tilde{\rho}(0)-i\alpha\int_{0}^{t}{dt}^{\prime}\left[\sum_{i=1}^{n}
	\tilde{V}_{ib_i}(t^{\prime})+\tilde{H}_{I}(t^{\prime}),\tilde{\rho}(t^{\prime
	})\right]. \label{chutiaojian}
\end{equation}
Then insert Eq. \eqref{chutiaojian} into Eq.~\eqref{interaction ev}, and keep terms up to the second order of the small parameter $\alpha$. For scenarios where the correlation time $\tau_b$ of the bath is small compared to the relaxation time of the open system, the bath will continuously lose the information of the connected system, and then the Markov approximation can be used to simplify the calculation.
As a result, the dependence on the past state $\tilde{\rho}(t')$ can be neglected, and one gets
\begin{eqnarray}\label{eq:SOVNEiIP}
\partial_{t}\tilde{\rho}(t)  &=&-i\alpha\left[\tilde{H}_{int}(t)  ,\tilde{\rho}(0)\right]\nonumber \\
& &-\alpha^{2}\int_{0}^{t}{dt^{\prime}}\left[ \tilde{H}_{int}(t),\left[  \tilde{H}_{int}(t'),\tilde{\rho}(t)\right]  \right],
\end{eqnarray}
where for brevity, we let $\tilde{H}_{int}(t)\equiv\sum_{i=1}^n\tilde
{V}_{ib_i}(t)+\tilde{H}_{I}(t)$.  Due to the weak coupling between the system and the bath, the influence of the system on the bath can be neglected, and this treatment is the so-called Born approximation. Thus, the state of the total system can be approximately written as $\tilde{\rho}(t)\approx \tilde{\rho}_s(t)\otimes\prod_i\tau_i$, where $\tilde{\rho}_s(t)$ is the reduced density matrix of the system and $\tau_i$ is the stationary state of the bath $b_i$. Note that the system-bath coupling $V_{ib_i}$ can be generally written as $V_{ib_i}=\sum_{\mu}A^\mu_i\otimes B^\mu_{i}$ \cite{open3,open4}, where $A_i^{\mu}$ and $B_i^{\mu}$ are Hermitian operators of the system $i$ and the bath $b_i$, respectively. Taking trace over the baths and assuming that $\tr(\tau_iB_i^\mu )=0$~\cite{footnote}, one arrives at
\begin{equation}\label{eq:SSOIDE}
\begin{split}
\partial_t\tilde{\rho}_s(t)=&-i\alpha\left[\tilde{H}_I(t),\tilde{\rho}_s(0)\right]\\
&-\alpha^2\int_0^tdt{^\prime}\left[\tilde{H}_I(t),\left[\tilde{H}_I(t^\prime),\tilde{\rho}_s(t)\right]\right]\\
&-\alpha^2\sum_i\int^t_0dt^{\prime}\tr_b\left[\tilde{V}_{ib_i}(t),\left[\tilde{V}_{ib_i}(t^\prime),\tilde{\rho}_s(t)\right]\right],
\end{split}
\end{equation}
where those cross terms, such as $\tr_b(\tilde{H}_I\tilde{V}_{ib_i}\tilde{\rho})$ and $\tr_b(\tilde{V}_{ib_i}\tilde{V}_{jb_j}\tilde{\rho})\,(i\neq j)$, vanish due to $\tr(\tau_iB_i^{\mu})=0$. 

Focusing on similar circumstances with the traditional GKLS master equation, we can further apply the secular approximation. The resulting equation of motion will be the generator of a completely positive and trace-preserving map.
To this end, we decompose those operators acting on the Hilbert space of the system $\mathbb{S}$ in the eigenbasis of $H_s\equiv\sum_{i=1}^{n}H_{i}$. For instance, the decomposition of the intersubsystem interaction $H_I$ can be expressed as
\begin{equation}  \label{decomp}
\tilde{H}_I\left(t\right)=e^{iH_{s}t}H_Ie^{-iH_{s}t}
    =\sum_{\omega}e^{-i\omega t}H_I\left(\omega\right),
\end{equation}
where $H_I(\omega)=\sum_{\varepsilon_{n}-\varepsilon_{m}=\omega}\Pi\left(\varepsilon_{m}\right)H_I\Pi\left(\varepsilon_{n}\right)$ with $\Pi(\varepsilon)$ being the projector on the eigenspace associated to the energy $\varepsilon$ of $H_s$. Substituting those interaction operators of the decomposition form into Eq.~\eqref{eq:SSOIDE} and doing the integration once again, one obtains
\begin{widetext}
\begin{eqnarray}\label{S EV}
    \tilde{\rho}_{s}(u)  &=&\tilde{\rho}_{s}(0)-i\alpha\int_{0}^{u}dt\sum_{\omega}e^{-i\omega t}\left[
    H_{I}(\omega),\tilde{\rho}_{s}(0)\right]\nonumber-\alpha^2\int_0^udt\sum_{\omega,\omega^{\prime}}e^{-i\omega t}\left[H_I(\omega),\left[H_I(\omega^{\prime}),\tilde{\rho}_s(t)\right]\right]\int_0^t dt^{\prime}e^{-i\omega^{\prime}t^{\prime}}\nonumber\\
    &  &+\alpha^{2}\sum\limits_{i=1}^n\int_{0}^{u}dt\sum\limits_{\omega
	,\omega^{\prime},\mu,\nu}e^{i(\omega-\omega^{\prime})t}\left[A_{i}^{\mu}(\omega)\tilde{\rho}_{s}(t)A_{i}^{{\nu}\dagger}(\omega^{\prime})-A_{i}^{\nu\dagger}(\omega^{\prime})A_{i}^{\mu}(\omega)\tilde{\rho}_{s}(t)\right]\Gamma_{i}^{\mu\nu}(\omega^{\prime})+H.c.
\end{eqnarray}
\end{widetext}
where $H.c.$ stands for the Hermitian conjugated expression, $A^{\mu\dagger}_i(\omega)=A^{\mu}_i(-\omega)$, and
$\Gamma_{i}^{{\mu\nu}}(\omega^{\prime})$ is the one-sided Fourier transformation of the  correlation function of the bath $b_i$, and is defined as
\begin{equation}		\Gamma_{i}^{{\mu\nu}}(\omega^{\prime})=\int_{0}^{\infty}dt^{\prime}e^{i\omega^{\prime}t^{\prime}}\tr\left[\tau_{{i}
	}  B_{i}^{\mu}(t^{\prime})B_{i}^{\nu}(0)\right],
\end{equation}
where we have taken the integral upper limit to infinity based on the fact that the correlation of the bath will decay rapidly~\cite{open3}. In order to reveal those dominant terms, following the standard procedure for deriving the GKLS form~\cite{open4}, we define rescaled times $\tau=\alpha t$, $\tau^{\prime}=\alpha t^{\prime}$, and $\sigma=\alpha u$. Then, Eq.~(\ref{S EV}) becomes
\begin{widetext}
\begin{eqnarray}\label{Beyond s-a}
	\tilde{\rho}_{s}(\frac{\sigma}{\alpha})  &=&\tilde{\rho}_{s}(0)-i\int_{0}^{\sigma}d\tau
	\sum_{\omega}e^{-i\frac{\omega}{\alpha}\tau}\left[H_{I}(\omega),\tilde{\rho}_{s}(0)\right]-\int_{0}^{\sigma}d\tau\sum\limits_{\omega,\omega^{\prime}}
	e^{-i\frac{\omega}{\alpha}\tau}\left[H_I(\omega),\left[H_I(\omega^{\prime}),\tilde{\rho}_s(\frac{\tau}{\alpha})\right]\right]\int_{0}^{\tau}{d\tau}^{\prime}e^{-i\frac{\omega^{\prime}	}{\alpha}\tau^{\prime}}\nonumber\\
	& &+\alpha\sum\limits_{i=1}^{n}\int_{0}^{\sigma}d\tau\sum\limits_{\omega,\omega^{\prime},\mu,\nu}e^{i\frac{\omega-\omega^{\prime}}{\alpha}\tau}\left[A_{i}^{\mu}(\omega)\tilde{\rho}_{s}(\frac{\tau}{\alpha})A_{i}^{{\nu}\dagger}(\omega^{\prime})-A_{i}^{\nu\dagger}(\omega^{\prime})A_{i}^{\mu}(\omega)\tilde{\rho}_{s}(\frac{\tau}{\alpha})\right]\Gamma_{i}^{\mu\nu}(\omega^{\prime})+h.c..
\end{eqnarray}
\end{widetext}
Taking the weak coupling limit $\alpha\rightarrow0$ while keeping $\sigma$ and $\tau$ finite, those terms satisfying $|\omega|\gg\mathcal{O}(\alpha)$, $|\omega^{\prime}|\gg\mathcal{O}(\alpha)$ and $|\omega-\omega^{\prime
}|\gg\mathcal{O}(\alpha)$ vanish due to the Riemann-Lebesgue lemma~\cite{open4}, which states that the integration of a highly oscillating function over a finite interval approaches zero.
At this stage, it is not enough to obtain the GKLS form. Thus, we introduce further assumptions: $|\omega|\gg\mathcal{O}(\alpha)$ and $|\omega-\omega^{\prime}|\gg\mathcal{O}(\alpha)$, which, roughly speaking, means our equation holds for scenarios where
the energy spectrum of $H_{s}$ is sparse. Then, in the first and second lines of Eq.~\eqref{Beyond s-a}, only the $\omega=0$ and $\omega^{\prime}=0$ contributions should be kept, and in the third line of Eq.~\eqref{Beyond s-a}, only the $\omega=\omega^{\prime}$ contributions should be kept. This treatment is usually known as the secular approximation~\cite{open3,open4}. Different from other derivations of GKLS-form master equations, here the secular approximation is used to treat not only the system-bath coupling but also the intersubsystem interaction.\\

Finally, returning back to the original time variable $(u,t,t^{\prime})$ and turning to the differential equation in the Schr\"odinger picture, one obtains (see Appendix~\ref{sec:details} for more details)
\begin{equation}\label{final-ME}
\begin{split}
	\partial_{t}\rho_{s}
	=-i\left[ H_s+\alpha^2H_{LS}+\alpha H_{I}(0),\rho_{s}\right]+\alpha^2\sum_{i=1}^{n}D_{i}[\rho_{s}],
\end{split}
\end{equation}
where $H_I(0)\equiv H_I(\omega=0)$, and
\begin{equation*}
\begin{split}
    H_{LS}&=\sum_{i,\omega,\mu,\nu}S_{i}^{\mu\nu}(\omega)A_{i}^{\mu}(\omega)A_{i}^{\nu\dagger}(\omega),\\
    D_i[\rho_s]&=\sum_{\omega ,\mu ,\nu } \gamma _{i}^{{\mu \nu }}(\omega )A_{i}^{{\mu }%
		}(\omega )\rho _{s}A_{i}^{{\nu }\dagger }(\omega )\\
	&\quad-\frac{1}{2}\sum_{\omega ,\mu ,\nu }\gamma _{i}^{{%
				\mu \nu }}(\omega )\left\{A_{i}^{{\nu }\dagger }(\omega )A_{i}^{\mu }(\omega
		),\rho _{s}\right\}
\end{split}
\end{equation*}
are the Lamb shift and the dissipation operator of the subsystem $s_i$, respectively. Here, we have let $\Gamma_{i}^{\mu\nu}
(\omega)=\frac{1}{2}\gamma_{i}^{{\mu\nu}}(\omega)+iS_{i}^{\mu\nu}(\omega)$. Equation~\eqref{final-ME} is the modified local master equation that we derived. One notes that this equation is still in the GKLS form and only the intersubsystem interaction is modified.
In the next part,
we will generally show that this equation satisfies the first law and second law of thermodynamics at the same time.

Here, we end this part
with some notes about the modified master equation. (\romannumeral1) In the derivation of our modified master equation, we apply similar approximations as one did in deriving the traditional GKLS master equation---the second-order approximation, the Born-Markov approximation, and the secular approximation. One key difference is that the secular approximation is also applied to modify the intersubsystem interaction here. Hence, in addition to $|\omega-\omega^{\prime}|\gg\mathcal{O}(\alpha)$, the secular approximation imposes another constraint for our equation, that is, those non-zero energy gaps of each subsystem should be larger than the coupling coefficient $\alpha$ [$|\omega|\gg\mathcal{O}(\alpha$)].
Since we apply similar approximations as the familiar GKLS master equation, it is reasonable that there are no contradictions between these approximations in our derivation, and our equation is valid for scenarios which satisfy the corresponding constraints imposed by those approximations. (\romannumeral2) In order to use the secular approximation, we decompose operators in the eigenbasis of $H_s$. This makes the sparse spectrum condition [$|\omega|\gg\mathcal{O}(\alpha)$ and $|\omega-\omega^{\prime}|\gg\mathcal{O}(\alpha$)] stricter, because intuitively, the spectrum of $H_s$ will be quite dense for large $n$ (the number of subsystems) even though the spectrum of each subsystem is sparse. However, the decomposition basis will actually reduce to the eigenbasis of the Hamiltonian of several subsystems, and the corresponding spectrum will be sparser. For example, the decomposition basis of $A_i^{\mu}$ in the system-bath coupling will reduce to the eigenbasis of $H_i$, and this also reflects the local feature of our equation. 
(\romannumeral3) In order to eliminate those terms that satisfy $|\omega|\gg\mathcal{O}(\alpha)$, $|\omega^{\prime}|\gg\mathcal{O}(\alpha)$ and $|\omega-\omega^{\prime
}|\gg\mathcal{O}(\alpha)$ in Eq.~\eqref{Beyond s-a}, we take the weak coupling limit $\alpha\rightarrow 0$ while keeping $\sigma$ finite. The finite $\sigma$ implies that $\mathcal{O}(\alpha)\ll\sigma\ll\mathcal{O}(1/\alpha)$, or else the Riemann-Lebesgue lemma can no-longer applied in the weak coupling limit (as the integration interval is not finite when we take $\alpha\rightarrow 0$). Therefore, the working time-scale of Eq.~\eqref{final-ME} should be $\max\left\{\mathcal{O}(1),\mathcal{O}(\tau_b)\right\}\ll t\ll \mathcal{O}(1/\alpha^2)$, where $\tau_b$ is provided by the Markov approximation. One should note that the traditional GKLS master equation has a similar working time-scale as well, which is also generated by the secular approximation and the Markov approximation.  (\romannumeral4) In Eq.~\eqref{final-ME}, we express the contribution of $H_I$ through the commutator $[\alpha H_I(0),\rho_s]$. Following the procedure in Eqs.~\eqref{interaction ev}--\eqref{eq:SOVNEiIP} and up to $\alpha^2$, it is equivalent with the contribution of $H_I$ in Eq.~\eqref{Beyond s-a} (modified by the secular approximation). As $H_I(0)=\sum_{\varepsilon}\Pi(\varepsilon)H_I\Pi(\varepsilon)$, we know that the effect of $H_I(0)$ is producing transitions between energy levels in the degenerate eigensubspace of $H_s$.
(\romannumeral5) In the weak coupling limit, the order of magnitude of the energy shift resulting from the Lamb shift $\alpha^2H_{LS}$ is $\mathcal{O}(\alpha^2)$ , while the sparse spectrum condition requires that the energy gap $\omega$ should satisfy $|\omega|\gg\mathcal{O}(\alpha)$. Therefore, the contribution of the Lamb shift to the energy level can be safely ignored for those cases to which our modified local master equation is applicable.

\section{Thermodynamical consistency}
We then show that our local master equation is simultaneously consistent with the first and second laws of thermodynamics . In the following discussions, the Lamb shift is ignored, as mentioned above.

One notes that after doing the secular approximation, the weak intersubsystem interaction produces transitions in the degenerate eigensubspace of $H_s$. In this sense, the local heat current should come from the transition between energy levels of $H_s$. Thus, it is natural to define the local heat current flowing from bath $b_i$ to the system as
\begin{eqnarray}\label{Q}
    \dot{Q}_i=\alpha^2\tr\left(H_sD_i[\rho_s(t)]\right).
\end{eqnarray}
This definition is also used in Refs.~\cite{linden,heat4}.
Since the transition in the degenerate eigensubspace does not cost extra energies, for a system without external power, the dynamical version of the first law can be expressed as
\begin{align}
	\dot{E_{s}}=\sum\limits_{i=1}^{n}\dot{Q_{i}},\label{first-law}
\end{align}
where
\begin{eqnarray}
    \dot{E}_s=\frac{d}{dt}\tr[H_s\rho_s(t)]
\end{eqnarray}
is the change rate of the internal energy. We now show that Eq.~\eqref{first-law} (the first law of thermodynamics) holds for our modified local master equation.

Using the local master equation (\ref{final-ME}) and ignoring the Lamb shift, the left-hand side of Eq.~(\ref{first-law}) will be
\begin{equation}
\begin{split}
\dot{E_{s}}
&=\tr\left(-iH_{s}\left[  H_{s}+\alpha H_{I}(0),\rho
	_{s}\right]  \right) +\alpha^2\sum\limits_{i=1}^{n}\tr\left(  H_{s}D_{i}\left[  \rho_{s}\right] \right)\\
    &=\sum\limits_{i=1}^{n}\dot{Q_{i}},
\end{split}
\end{equation}
which proves Eq.~\eqref{first-law}. Note that in the third equality, we have used the equation $\tr\left(  H_{s}\left[  H_{s}+\alpha H_{I}(0),\rho
	_{s}\right]  \right)=0$.

For the entire setup that contains the baths and the subsystems, the second law of thermodynamics can be described by the irreversible entropy production:
\begin{equation}
	\frac{dS}{dt}-\sum\limits_{i=1}^{n}\beta_{i}\dot{Q_{i}}\geq0,
	\label{sahng}
\end{equation}
where $\beta_{{i}}$ denotes the inverse temperature of the bath $b_{i}$, and $S=-\tr(\rho_{s}\ln\rho_{s})$~\cite{nielsen}
is the entropy of the system $\mathbb{S}$. In the following, we will show how to produce Eq. \eqref{sahng} from our modified local master equation. The derivations are similar to those in Ref. \cite{heat6PRR}.

According to the definition of the entropy $S$ and the local master equation \eqref{final-ME}, the time derivative of $S$ can be expressed as
\begin{align}
	\frac{dS}{dt}=-\tr\left(  \mathcal{L}[\rho_{s}]\ln\rho_{s}\right),
\end{align}
where $\mathcal{L}$ is the Liouvillian superoperator and defined as
\begin{equation}\label{key}
	\mathcal{L}[\rho_s]=-i\left[ H_{s}
	+\alpha H_{I}(0),\rho_{s}\right]  +\alpha^2\sum\limits_{i=1}^{n}D_{i}\left[\rho_{s}\right].
\end{equation}
We further introduce a partial superoperator $\mathcal{L}^{\prime}$,
\begin{equation}
	\mathcal{L}^{\prime}[\rho_s]=-i\left[H_{s},\rho_{s}\right]
	+\alpha^2\sum\limits_{i=1}^{n}D_{{i}}[\rho_{s}].
\end{equation}
Note that $\tau
_{s}=\prod_{i}\exp\left(-\beta_{i}H_{i}\right)/Z$, with $Z$ being the partition function, is the steady state of the partial superoperator $\mathcal{L}^{\prime}$, that is,  $\mathcal{L}^{\prime}\left[\tau_s\right]=0$.
Applying the Spohn's inequality~\cite{spohn} to the partial superoperator $\mathcal{L}^{\prime}$, one gets
\begin{equation}\label{spohn}
	-\tr(\mathcal{L}^{\prime}[\rho_{s}]\ln\rho_{s})\geq-\tr(\mathcal{L}^{\prime}[\rho_{s}]\ln\tau_{s}).
\end{equation}
Inserting the explicit expression of $\tau_{s}$ into the right-hand side of Eq.~\eqref{spohn}, one has
\begin{equation}\label{eq:RHSoSI}
\begin{split}
-\tr(\mathcal{L}^{\prime}[\rho_{s}]\ln\tau_{s})&
=\alpha^2\sum_{i=1}^{n}\beta_{i}\tr\left(H_{s}D_{{i}}[\rho_{s}]\right)\\
&=\sum_{i=1}^{n}\beta_i\dot{Q}_i.
\end{split}
\end{equation}
The left-hand side of Eq.~\eqref{spohn} can be rewritten as
\begin{equation}\label{eq:LFSoSI}
\begin{split}
    &-\tr(\mathcal{L}^{\prime}[\rho_{s}]\ln\rho_{s})\\
    =&-\tr(\mathcal{L}^{\prime}[\rho_{s}]\ln\rho_{s})+i\tr\left\{[\alpha H_I(0),\rho_s]\ln\rho_s\right\}\\
    =&-\tr\left(\mathcal{L}[\rho_s]\ln\rho_s\right)\\
    =&\frac{dS}{dt},
\end{split}
\end{equation}
where, in the second equality, we have used the identity $\tr\left([H_I(0),\rho_s]\ln\rho_s\right)=0$. Combining Eqs.~\eqref{spohn}--\eqref{eq:LFSoSI}, the second law of thermodynamics, given by Eq.~\eqref{sahng}, can be obtained. And thus, we recover the thermodynamical consistency in our local master equation.

\section{Two-qubit heat transfer model}
As an example of our local master equation,
we consider a two-qubit heat transfer network~\cite{heat2021,glocal5,glocal0}. This model contains two qubits, each of which couples to a large bosonic bath with temperature $T_i\,(i=1,2)$. The total Hamiltonian reads
\begin{eqnarray}
H=H_s+ \alpha H_I+\sum_{i=1}^{2}(H_{b_i}+\alpha V_{ib_i}),
\end{eqnarray}
where $H_s=(E_1\sigma_1^z+E_2\sigma_2^z)/2$ is the bare Hamiltonian of the two qubits, $\alpha H_I=\alpha\sigma^x_1\sigma^x_2$ is the weak intersubsystem interaction,  $H_{b_i}=\int_{0}^{\infty}d\omega\,\omega a^{\dagger}_i(\omega)a_i(\omega)$ is the Hamiltonian of large baths $b_i$, and $\alpha V_{ib_i}=\int_{0}^{\infty}d\omega \alpha h_i(\omega)\sigma_i^x[a^{\dagger}_i(\omega)+a_i(\omega)]$ is the weak coupling between qubit $i$ and bath $b_i$. $\omega$ is the energy of bosonic modes in the baths, $\alpha h_i(\omega)$ is the coupling function, and $a_i(\omega)$ denotes the bosonic annihilation operator. For energies of the two qubits, $E_1$ and $E_2$, one can always tune them into the regime satisfying the sparse spectrum condition. Therefore, this example is a suitable platform to apply the second-order approximation, the Born-Markov approximation, and the secular approximation, and thus our equation can be used to treat this model.

For the case that $E_1=E_2\equiv E$, $\{\ket{10},\ket{01}\}$ in the basis of $\sigma^z$ is a degenerate eigensubspace of the Hamiltonian $H_s$. Thus, the intersubsystem interaction $H_I$ will be modified to be
\begin{eqnarray}\label{HI}
H_{I}(0)
&=&\sigma_{1}^{+}\otimes\sigma_{2}^-+\sigma_{1}^-\otimes\sigma
_{2}^{+},
\end{eqnarray}
where $\sigma^{\pm}=(\sigma^x\pm i\sigma^y)/2$.

Here, we study the behavior of the steady state in this model through our modified local master equation. Note that in the steady state, Eq.~\eqref{final-ME} becomes
\begin{equation}\label{steady}
	0=-i\left[ H_{s}+H_{I}(0),\rho_{ss}\right]
	+\alpha^2\sum\limits_{i=1}^{2}D_{i}[\rho_{ss}],
\end{equation}
where $\rho_{ss}$ is the density matrix of the steady state, the Lamb shift has been ignored, the condition that $\tr(\tau_iB_i^{\mu})=0$ is satisfied, and $D_{i}[\rho_{ss}]$ reads
\begin{equation}\label{disspator}
\begin{split}
&D_{i}[\rho_{ss}]\\ =&\,2\pi h_i^2(E_i)\left[n_{i}(E_{i})+1\right] \left[ \sigma_{i}^{-}\rho_{ss}
\sigma_{i}^{+}-\frac{1}{2}\{\sigma_{i}^{+}\sigma_{i}^{-},\rho_{ss} \}\right]  \\
&+2\pi h_i^2(E_i)n_i(E_{i})\left[\sigma_{i}^{+}\rho_{ss}\sigma_{i}^{-}-\frac{1}{2}\{\sigma_{i}^{-}\sigma_{i}^{+},\rho_{ss} \}\right],
\end{split}
\end{equation}
where $n_i(\omega)=1/(e^{\beta_{i}\omega}-1)$ is the Bose-Einstein distribution of the bosonic bath.

Combining Eqs.~\eqref{steady} and \eqref{disspator}, the steady state can be obtained. And then using Eq.~\eqref{Q}, one can get the heat current flowing from the bath $b_1$ to the qubits, that is,
\begin{eqnarray}
\dot{Q}_{1}&=&\frac{e^{\beta
_{2}E}-e^{\beta_{1}E}}{\left( e^{\beta_{1}E}+1\right) \left(
e^{\beta_{2}E}+1\right) }\mathcal{F},
\end{eqnarray}
where $\mathcal{F}$ is a positive function. The heat current flowing from bath $b_2$ to the qubits can be derived in a similar way and can be expressed as $\dot{Q}_2=-\dot{Q}_1$. Since $\dot{Q}_1+\dot{Q}_2=0$ and the change rate of the internal energy is zero, it is obvious that the first law of thermodynamics is satisfied. One also finds that when $\beta_1<\beta _2$, $\dot{Q}_1>0$ and $\dot{Q}_2<0$, which means that the heat current flows from the hot bath to the system, and then to the cold bath. This process conforms to the second law of thermodynamics.

As for $E_{1}\neq E_{2}$, the intersubsystem interaction dose not contribute to the local master equation, and then it reduces to two separated local master equations -- one for the qubit $1$ and the bath $b_1$ and one for the qubit $2$ and the bath $b_2$. In the steady state, $\dot{Q}_1=\dot{Q}_2=0$, which means no heat current exists between these two baths. This case is also consistent with the second law of thermodynamics, as now two baths are disconnected.

This result can be interpreted through the following picture. In the local approach, subsystems are localized and do not form collective modes, thus the weak intersubsystem interaction can be regarded as external operations but without implementers. As implementers do not exist, those fake external operations can be implemented only if they do not cost extra work. This is similar to the idea of a self-contained refrigerator (not requiring external sources of work)~\cite{linden}. Therefore, only $H_I(0)$, which produces transitions in the degenerate subspace, will contribute to the evolution. Since now those unphysical processes encoded in $H_I(\omega\neq 0)$ are excluded, it is obvious that the second law of thermodynamics will not be violated.

\section{Conclusion}
In this work, we have presented an approach to derive an alternative thermodynamically-consistent local master equation, which works when the spectrum condition and the time scale condition are satisfied.
Based on a concise and natural definition of the heat current, we generally prove that the modified local master equation fulfills the first and second laws of thermodynamics at the same time. This good property suggests that our modified local master equation may be a more suitable method to deal with weak-interacting systems, especially when the thermodynamic properties of open quantum systems are considered.

 \begin{acknowledgments}
This work was supported by the National Natural Science
Foundation of China (Grants No. 11675119, No. 11575125, and No. 11105097).
 \end{acknowledgments}
 
\appendix
\begin{widetext}
\section{Details about the derivation of the modified local master equation }\label{sec:details}
In this section, we show more details about the derivation of our modified local master equation and discuss the case in which strengths of the system-bath coupling and the intersubsystem interaction are different.

Here, we consider a more general time-independent Hamiltonian of the total system,
\begin{equation}
	H=\sum_{i=1}^{n}(H_{i}+H_{b_i}+\beta V_{ib_i})+\alpha H_{I},	
\end{equation}
where $\alpha$ and $\beta$ are different interaction strength, but both in the weak coupling regime. In the interaction picture, after doing the second-order approximation with respect to $\alpha,\beta$ and the Born-Markov approximation as we discussed in the main text, one arrives at Eq. (6) in the main text, and we also show it here:
\begin{equation}\label{eq:SSOIDE2}
\partial_t\tilde{\rho}_s(t)=-i\alpha\left[\tilde{H}_I(t),\tilde{\rho}_s(0)\right]-\alpha^2\int_0^tdt{^\prime}\left[\tilde{H}_I(t),\left[\tilde{H}_I(t^\prime),\tilde{\rho}_s(t)\right]\right]-\beta^2\sum_i\int^t_0dt^{\prime}\tr_b\left[\tilde{V}_{ib_i}(t),\left[\tilde{V}_{ib_i}(t^\prime),\tilde{\rho}_s(t)\right]\right].
\end{equation}
In the derivation, one will encounter cross terms, such as $\tr_b(\tilde{H}_I\tilde{V}_{ib_i}\tilde{\rho})$ and $\tr_b(\tilde{V}_{ib_i}\tilde{V}_{jb_j}\tilde{\rho})\,(i\neq j)$. Here we briefly show how they vanish due to $\tr(\tau_iB_i^{\mu})=0$,
\begin{equation}
\begin{split}
\tr_b\left[\tilde{V}_{ib_i}(t)\tilde{H}_I(t^{\prime})\tilde{\rho}(t)\right]
&=\tr_{b}\left[\sum_{\mu}\tilde{A}_{i}^{\mu}\left(t\right)\otimes\tilde{B}_{i}^{\mu}\left(t\right)\tilde{H}_I(t^{\prime})\tilde{\rho}_s(t)\otimes\prod_j\tau_j\right]\\
&=\sum_{\mu}\tilde{A}_{i}^{\mu}\left(t\right)\tilde{H}_I(t^{\prime})\tilde{\rho}_s(t)\cdot\prod_{j\neq i}\tr_{b_j}\left(\tau_j\right)\cdot\tr_{b_i}\left[\tau_i\tilde{B}^{\mu}_i(t)\right]\\
&=\sum_{\mu}\tilde{A}_{i}^{\mu}\left(t\right)\tilde{H}_I(t^{\prime})\tilde{\rho}_s(t)\cdot\prod_{j\neq i}\tr_{b_j}\left(\tau_j\right)\cdot\tr_{b_i}\left[\tau_iB^{\mu}_i\right]\\
&=0,
\end{split}
\end{equation}
where the third equality is due to the fact that $\tau_i$ is a stationary state of the bath $b_i$, that is, $[H_{b_i},\tau_i]=0$. Other cross terms can be calculated in a similar way.

To ensure that the equation of motion is the generator of a completely positive and trace-preserving map, we further apply the secular approximation. To this end, we decompose those operators acting on the Hilbert space of the system $\mathbb{S}$ in the eigenbasis of $H_s\equiv\sum_{i=1}^{n}H_{i}$ [see Eq.~\eqref{decomp}]. One notes that the decomposition basis, which is formed by the eigenstates of $H_s$, will reduce to eigenstates of each subsystem, that is, eigenstates of $H_i$. Thus, our master equation is indeed a local approach.

Substituting those interaction operators of the decomposition form into Eq.~\eqref{eq:SSOIDE2} and doing the integration once again, one obtains
\begin{eqnarray}\label{S EV2}
    \tilde{\rho}_{s}(u)  &=&\tilde{\rho}_{s}(0)-i\alpha\int_{0}^{u}dt\sum_{\omega}e^{-i\omega t}\left[
    H_{I}(\omega),\tilde{\rho}_{s}(0)\right]-\alpha^2\int_0^udt\sum_{\omega,\omega^{\prime}}e^{-i\omega t}\left[H_I(\omega),\left[H_I(\omega^{\prime}),\tilde{\rho}_s(t)\right]\right]\int_0^t dt^{\prime}e^{-i\omega^{\prime}t^{\prime}}\nonumber\\
    &  &+\beta^{2}\sum\limits_{i=1}^n\int_{0}^{u}dt\sum\limits_{\omega
	,\omega^{\prime},\mu,\nu}e^{i(\omega-\omega^{\prime})t}\left[A_{i}^{\mu}(\omega)\tilde{\rho}_{s}(t)A_{i}^{{\nu}\dagger}(\omega^{\prime})-A_{i}^{\nu\dagger}(\omega^{\prime})A_{i}^{\mu}(\omega)\tilde{\rho}_{s}(t)\right]\Gamma_{i}^{\mu\nu}(\omega^{\prime})+H.c.
\end{eqnarray}
Following the standard procedure for deriving the GKLS form~\cite{open4}, we define rescaled times $\tau=\alpha t$, $\tau^{\prime}=\alpha t^{\prime}$, and $\sigma=\alpha u$. We have assumed that $\alpha>\beta$ and one can treat the $\alpha<\beta$ case similarly. Then, Eq.~(\ref{S EV2}) becomes
\begin{eqnarray}\label{Beyond s-a2}
	\tilde{\rho}_{s}(\frac{\sigma}{\alpha})  &=&\tilde{\rho}_{s}(0)-i\int_{0}^{\sigma}d\tau
	\sum_{\omega}e^{-i\frac{\omega}{\alpha}\tau}\left[H_{I}(\omega),\tilde{\rho}_{s}(0)\right]-\int_{0}^{\sigma}d\tau\sum\limits_{\omega,\omega^{\prime}}
	e^{-i\frac{\omega}{\alpha}\tau}\left[H_I(\omega),\left[H_I(\omega^{\prime}),\tilde{\rho}_s(\frac{\tau}{\alpha})\right]\right]\int_{0}^{\tau}{d\tau}^{\prime}e^{-i\frac{\omega^{\prime}	}{\alpha}\tau^{\prime}}\nonumber\\
	& &+\frac{\beta^2}{\alpha}\sum\limits_{i=1}^{n}\int_{0}^{\sigma}d\tau\sum\limits_{\omega,\omega^{\prime},\mu,\nu}e^{i\frac{\omega-\omega^{\prime}}{\alpha}\tau}\left[A_{i}^{\mu}(\omega)\tilde{\rho}_{s}(\frac{\tau}{\alpha})A_{i}^{{\nu}\dagger}(\omega^{\prime})-A_{i}^{\nu\dagger}(\omega^{\prime})A_{i}^{\mu}(\omega)\tilde{\rho}_{s}(\frac{\tau}{\alpha})\right]\Gamma_{i}^{\mu\nu}(\omega^{\prime})+H.c..
\end{eqnarray}
Taking the weak coupling limit $\alpha\rightarrow0$ while keeping $\sigma$ and $\tau$ finite, those terms satisfying $|\omega|\gg\mathcal{O}(\alpha)$, $|\omega^{\prime}|\gg\mathcal{O}(\alpha)$ and $|\omega-\omega^{\prime
}|\gg\mathcal{O}(\alpha)$, vanish due to the Riemann-Lebesgue lemma \cite{open4}:

{\textbf{Lemma}}$\,\,\,${\em Let $f(t)$ be integrable in a finite interval $[a,b]$, then}
\begin{equation*}
    \lim_{x\rightarrow\infty}\int_a^bdt\,e^{ixt}f(t)=0.
\end{equation*}
In the main text, we introduce further assumptions: $|\omega|\gg\mathcal{O}(\alpha)$ and $|\omega-\omega^{\prime}|\gg\mathcal{O}(\alpha)$. Then, in the first and second lines of Eq.~\eqref{Beyond s-a2}, only the $\omega=0$ and $\omega^{\prime}=0$ contributions should be kept, and in the third line of Eq.~\eqref{Beyond s-a2}, only the $\omega=\omega^{\prime}$ contributions should be kept.

After the secular approximation, one arrives at
\begin{equation}
\begin{split}
 \tilde{\rho}_{s}(\frac{\sigma}{\alpha})  &=\tilde{\rho}_{s}(0)-i\int_{0}^{\sigma}d\tau
	\left[H_{I}(0),\tilde{\rho}_{s}(0)\right]-\int_{0}^{\sigma}d\tau
	\left[H_I(0),\left[H_I(0),\tilde{\rho}_s(\frac{\tau}{\alpha})\right]\right]\int_{0}^{\tau}{d\tau}^{\prime}\cdot1\\
	&\quad+\frac{\beta^2}{\alpha}\sum\limits_{i=1}^{n}\int_{0}^{\sigma}d\tau\sum\limits_{\omega,\mu,\nu}\left[A_{i}^{\mu}(\omega)\tilde{\rho}_{s}(\frac{\tau}{\alpha})A_{i}^{{\nu}\dagger}(\omega)-A_{i}^{\nu\dagger}(\omega)A_{i}^{\mu}(\omega)\tilde{\rho}_{s}(\frac{\tau}{\alpha})\right]\Gamma_{i}^{\mu\nu}(\omega)+H.c..
\end{split}
\end{equation}
By defining rescaled times, we connect the energy gaps with the coupling strength, and reveal the dominant terms for the evolution. This treatment is analogous to the regularization procedure in eliminating infinity in quantum field theory. Returning back to the original times, we have
\begin{equation}
\begin{split}\label{eq:rhos}
 \tilde{\rho}_{s}(u)  &=\tilde{\rho}_{s}(0)-i\alpha u
	\left[H_{I}(0),\tilde{\rho}_{s}(0)\right]-\alpha^2\int_{0}^{u}dtt
	\left[H_I(0),\left[H_I(0),\tilde{\rho}_s(t)\right]\right]\\
	&\quad+\beta^2\sum\limits_{i=1}^{n}\int_{0}^{u}dt\sum\limits_{\omega,\mu,\nu}\left[A_{i}^{\mu}(\omega)\tilde{\rho}_{s}(t)A_{i}^{{\nu}\dagger}(\omega)-A_{i}^{\nu\dagger}(\omega)A_{i}^{\mu}(\omega)\tilde{\rho}_{s}(t)\right]\Gamma_{i}^{\mu\nu}(\omega)+h.c..
\end{split}
\end{equation}
Rewriting the above equation in the differential form, we have
\begin{equation}
    \begin{split}\label{eq:bfe}
\partial_t\tilde{\rho}_s(t)&=-i\alpha\left[H_I(0),\tilde{\rho}_s(0)\right]-\alpha^2t\left[H_I(0),\left[H_I(0),\tilde{\rho}_s(t)\right]\right]\\
&\quad+\beta^2\sum_{i=1}^{n}\sum_{\omega,\mu,\nu}\left[A_{i}^{\mu}(\omega)\tilde{\rho}_{s}(t)A_{i}^{{\nu}\dagger}(\omega)-A_{i}^{\nu\dagger}(\omega)A_{i}^{\mu}(\omega)\tilde{\rho}_{s}(t)\right]\Gamma_{i}^{\mu\nu}(\omega)+H.c.\\
&=-i\alpha\left[H_I(0),\tilde{\rho}_s(0)\right]-\alpha^2\int_0^tdt_1\left[H_I(0),\left[H_I(0),\tilde{\rho}_s(t)\right]\right]\\
&\quad+\beta^2\sum_{i=1}^{n}\sum_{\omega,\mu,\nu}\left[A_{i}^{\mu}(\omega)\tilde{\rho}_{s}(t)A_{i}^{{\nu}\dagger}(\omega)-A_{i}^{\nu\dagger}(\omega)A_{i}^{\mu}(\omega)\tilde{\rho}_{s}(t)\right]\Gamma_{i}^{\mu\nu}(\omega)+H.c..
    \end{split}
\end{equation}
This equation can be further approximated as
\begin{equation}
\begin{split}\label{eq:feiip}
 \partial_t\tilde{\rho}_s(t)= -i\alpha\left[H_I(0),\tilde{\rho}_s(t)\right]+\beta^2\sum_{i=1}^{n}\sum_{\omega,\mu,\nu}\left[A_{i}^{\mu}(\omega)\tilde{\rho}_{s}(t)A_{i}^{{\nu}\dagger}(\omega)-A_{i}^{\nu\dagger}(\omega)A_{i}^{\mu}(\omega)\tilde{\rho}_{s}(t)\right]\Gamma_{i}^{\mu\nu}(\omega)+H.c..
\end{split}
\end{equation}

The error of this treatment is about $\mathcal{O}(\alpha^3)$. As the leading order of $H_I$ is $\mathcal{O}(\alpha)$, $\mathcal{O}(\alpha^3)$ errors have very limited contributions.

We now prove that Eq. \eqref{eq:feiip} differs from Eq. \eqref{eq:bfe} up to $\mathcal{O}(\alpha^3)$. Integrating Eq. \eqref{eq:feiip} once, one gets
\begin{equation}
\begin{split}
    \tilde{\rho}(t)=\tilde{\rho}_s(0)-i\alpha\int_0^t dt^{\prime}\left[H_I(0),\tilde{\rho}_s(t^{\prime})\right]+\beta^2\sum_{i=1}^{n}\sum_{\omega,\mu,\nu}\int_0^{t^{\prime}}dt^{\prime}\left[A_{i}^{\mu}(\omega)\tilde{\rho}_{s}(t^{\prime})A_{i}^{{\nu}\dagger}(\omega)-A_{i}^{\nu\dagger}(\omega)A_{i}^{\mu}(\omega)\tilde{\rho}_{s}(t^{\prime})\right]\Gamma_{i}^{\mu\nu}(\omega)+H.c..
    \end{split}
\end{equation}
Substituting this into $[H_I(0),\tilde{\rho}_s(t)]$ in Eq. \eqref{eq:feiip} and keeping terms up to $\alpha^2$, one has
\begin{equation}
 \begin{split}   \partial_t\tilde{\rho}_s(t)&=-i\alpha\left[H_I(0),\tilde{\rho}_s(0)\right]-\alpha^2\int_0^t dt^{\prime}\left[H_I(0),\left[H_I(0),\tilde{\rho}_s(t^{\prime})\right]\right]\\
 &\quad+\beta^2\sum_{i=1}^{n}\sum_{\omega,\mu,\nu}\left[A_{i}^{\mu}(\omega)\tilde{\rho}_{s}(t)A_{i}^{{\nu}\dagger}(\omega)-A_{i}^{\nu\dagger}(\omega)A_{i}^{\mu}(\omega)\tilde{\rho}_{s}(t)\right]\Gamma_{i}^{\mu\nu}(\omega)+H.c..
 \end{split}
\end{equation}
The Markov approximation thus leads to Eq. \eqref{eq:bfe}.

Finally, letting $\Gamma_{i}^{\mu\nu}
(\omega)=\frac{1}{2}\gamma_{i}^{{\mu\nu}}(\omega)+iS_{i}^{\mu\nu}(\omega)$ and then returning to the Schr\"odinger picture, Eq. \eqref{eq:feiip} becomes:
\begin{equation}\label{final-ME2}
\begin{split}
	\partial_{t}\rho_{s}
	=-i\left[ H_s+\beta^2H_{LS}+\alpha H_{I}(0),\rho_{s}\right]+\beta^2\sum_{i=1}^{n}D_{i}[\rho_{s}].
\end{split}
\end{equation}
We thus obtain the modified local master equation for $\alpha\neq\beta$, and it is in the same form as Eq. (11) in the main text. \\
\section{Steady state of the partial Liouvillian superoperator}\label{sec:proof}
In this section, we show that $\tau _{s}=\prod_{i}\exp \left( -\beta _{i}H_{i}\right) /Z$ is the steady state of the partial Liouvillian superoperator $\mathcal{L}^{\prime}$.  Before rigorously proving this statement, we first provide a physical interpretation. When the system is in contact with a bath, temperatures of the system and the bath will reach the same value. At this time, the steady state of the system will be the thermal equilibrium state. This is similar to the derivation of the canonical ensemble in statistical mechanics.

We now provide a brief proof to illustrate this. In order to show that $\tau _{s}=\prod_{i}\exp \left( -\beta _{i}H_{i}\right) /Z$ is the steady state of $\mathcal{L}^{\prime}$, we need to show $\mathcal{L^{\prime }}(\tau _{s})=0$. Since $\tau_s$ commutes with $H_{s}=\sum_iH_i$, that is $\left[ H_{s},\tau _{s}\right] =0$, we just need to prove
\begin{equation}
		\sum_{\omega ,\mu ,\nu }\gamma _{i}^{{\mu \nu }}(\omega )\left[A_{i}^{{\mu }%
		}(\omega )\tau _{s}A_{i}^{{\nu }\dagger }(\omega )-\frac{1}{2}\left\{A_{i}^{{\nu }\dagger }(\omega )A_{i}^{\mu }(\omega
		),\tau _{s}\right\}\right] =0.  \label{1}
	\end{equation}
Since $\omega $ and $-\omega $ appear in pairs, Eq. \eqref{1} then reduces to
\begin{equation}
		\sum_{\omega ,\mu ,\nu }\left[\gamma _{i}^{{\mu \nu }}(\omega )A_{i}^{{\mu }%
		}(\omega )\tau _{s}A_{i}^{{\nu }\dagger }(\omega )-\frac{1}{2}\gamma _{i}^{{\mu \nu }}(-\omega)\left\{A_{i}^{{\nu }\dagger }(-\omega)A_{i}^{\mu }(-\omega),\tau _{s}\right\}\right] =0.  \label{eq:dissipator}
	\end{equation}
Given that $\tau_s=\prod_i\exp(-\beta_iH_i)/Z,$ Eq.~\eqref{eq:dissipator} can further reduce to
\begin{equation}
		\sum_{\omega ,\mu ,\nu }\left[\gamma _{i}^{{\mu \nu }}(\omega )A_{i}^{{\mu }%
		}(\omega )e^{-\beta_iH_i}A_{i}^{{\nu }\dagger }(\omega )-\frac{1}{2}\gamma _{i}^{{\mu \nu }}(-\omega)\left\{A_{i}^{{\nu }\dagger }(-\omega)A_{i}^{\mu }(-\omega),e^{-\beta_iH_i}\right\}\right] =0.  \label{eq:dissipator2}
	\end{equation}
Substituting $A_{i}^{{\mu }}(\omega )=\sum_{\varepsilon _{n}-\varepsilon
		_{m}=\omega }\Pi \left( \varepsilon _{m}\right) A_{i}^{{\mu }}\Pi \left(
	\varepsilon _{n}\right)$ into the first
term, one will get
\begin{eqnarray}
&&\gamma _{i}^{{\mu \nu }}(\omega )A_{i}^{{\mu }}(\omega )e^{-\beta_iH_i}A_{i}^{{%
\nu }\dagger }(\omega )  \label{2} \\
&=&\gamma _{i}^{{\mu \nu }}(\omega )\sum_{\varepsilon _{n}-\varepsilon
_{m}=\omega }\Pi \left( \varepsilon _{m}\right) A_{i}^{{\mu }}\Pi \left(
\varepsilon _{n}\right) e^{-\beta_iH_i}\left( \sum_{\varepsilon _{n}^{\prime
}-\varepsilon _{m}^{\prime }=\omega }\Pi \left( \varepsilon ^{\prime }_{n}\right)
A_{i}^{{\nu}}\Pi \left( \varepsilon^{\prime } _{m}\right) \right)   \nonumber\\
&=&\gamma _{i}^{{\nu \mu }}(-\omega )e^{\beta_i \omega }\sum_{\varepsilon
_{n}-\varepsilon _{m}=\omega }\sum_{\varepsilon^{\prime}_{n}-\varepsilon
_{m}^{\prime }=\omega }\Pi \left( \varepsilon _{m}\right) A_{i}^{{\mu }}\Pi
\left( \varepsilon _{n}\right) e^{-\beta_iH_i}\Pi \left(\varepsilon^{\prime}_{n}\right)
A_{i}^{{\nu}}\Pi \left( \varepsilon _{m}^{\prime }\right)   \notag \\
&=&\gamma _{i}^{{\nu \mu }}(-\omega )e^{\beta _{i}\omega }\sum_{\varepsilon
_{n}-\varepsilon _{m}=\omega }\sum_{\varepsilon _{n}-\varepsilon
_{m}^{\prime }=\omega }e^{-\beta_i\varepsilon_n}\Pi
\left( \varepsilon _{m}\right) A_{i}^{{\mu }}\Pi \left( \varepsilon
_{n}\right) \Pi \left( \varepsilon _{n}\right) A_{i}^{{\nu }}\Pi \left(
\varepsilon _{m}^{\prime }\right),   \nonumber
\end{eqnarray}
where the second equality is due to the detailed balance condition: $\gamma _{i}^{{\mu \nu }%
}(\omega )=\gamma _{i}^{{\nu \mu }}(-\omega )e^{\beta _{i}\omega }$~\cite{open3,pottier}. Similarly, the second term in Eq. \eqref{eq:dissipator2} becomes
\begin{eqnarray}
&&\frac{1}{2}\gamma _{i}^{{\mu \nu }}(-\omega )\left\{A_{i}^{{\nu }\dagger
}(-\omega )A_{i}^{\mu }(-\omega ),e^{-\beta_iH_i}\right\}  \label{3} \\
&=&\gamma _{i}^{{\mu \nu }}(-\omega )\sum_{\varepsilon _{n}-\varepsilon
_{m}=\omega }\sum_{\varepsilon _{n}-\varepsilon
_{m}^{\prime}=\omega }e^{-\beta
_{i}(\varepsilon _{n}-\omega )}\Pi \left( \varepsilon _{m}\right) A_{i}^{{\nu}}\Pi
\left( \varepsilon _{n}\right) \Pi \left( \varepsilon _{n}\right) A_{i}^{{%
\mu }}\Pi \left(\varepsilon _{m}^{\prime }\right).  \nonumber
\end{eqnarray}
Together with Eq. \eqref{2}, we have
\begin{equation}
\begin{split}
&\quad\sum_{\mu ,\nu }\gamma _{i}^{{\mu \nu }}(\omega )A_{i}^{{\mu }}(\omega)e^{-\beta_iH_i}A_{i}^{{\nu }\dagger }(\omega )-\frac{1}{2}\sum_{\mu ,\nu }\gamma
_{i}^{{\mu \nu }}(-\omega )\left\{A_{i}^{{\nu }\dagger }(-\omega )A_{i}^{\mu
}(-\omega ),e^{-\beta_iH_i}\right\} \\
&=\sum_{\mu,\nu}\gamma _{i}^{{\nu \mu }}(-\omega )e^{\beta _{i}\omega }\sum_{\varepsilon
_{n}-\varepsilon _{m}=\omega }\sum_{\varepsilon _{n}-\varepsilon
_{m}^{\prime }=\omega }e^{-\beta_i\varepsilon_n}\Pi
\left( \varepsilon _{m}\right) A_{i}^{{\mu }}\Pi \left( \varepsilon
_{n}\right) \Pi \left( \varepsilon _{n}\right) A_{i}^{{\nu }}\Pi \left(
\varepsilon _{m}^{\prime }\right)\\
&\quad-\sum_{\mu,\nu}\gamma _{i}^{{\mu \nu }}(-\omega )\sum_{\varepsilon _{n}-\varepsilon
_{m}=\omega }\sum_{\varepsilon _{n}-\varepsilon
_{m}^{\prime}=\omega }e^{-\beta
_{i}(\varepsilon _{n}-\omega )}\Pi \left( \varepsilon _{m}\right) A_{i}^{{\nu}}\Pi
\left( \varepsilon _{n}\right) \Pi \left( \varepsilon _{n}\right) A_{i}^{{%
\mu }}\Pi \left(\varepsilon _{m}^{\prime }\right)\\
&=\sum_{\mu ,\nu }\sum_{\varepsilon
_{n}-\varepsilon _{m}=\omega }\sum_{\varepsilon _{n}-\varepsilon
_{m}^{\prime }=\omega }\gamma _{i}^{{\mu \nu }}(-\omega )e^{-\beta _{i}(\varepsilon _{n}-\omega )}
\left[ \Pi \left( \varepsilon _{m}\right) A_{i}^{{\nu}}\Pi \left(
\varepsilon _{n}\right) A_{i}^{{\mu }}\Pi \left( \varepsilon _{m}^{\prime
}\right) -\Pi \left( \varepsilon _{m}\right) A_{i}^{{\nu}}\Pi \left(
\varepsilon _{n}\right) A_{i}^{{\mu }}\Pi \left( \varepsilon _{m}^{\prime}\right) \right]\\
&=0.
\end{split}
\end{equation}
Therefore, Eq.~\eqref{eq:dissipator2} and then Eq.~\eqref{eq:dissipator} can be proved.
\end{widetext}
\bibliographystyle{unsrt}
\bibliography{MLME_Ref}

\begin{thebibliography}{10}
\bibitem{footnote}
If this assumption is not satisfied, one can define a new interaction
  Hamiltonian between the subsystem and the attached bath, and get a shifted
  Hamiltonian of the subsystem. For example, if
  $\operatorname{Tr}(\tau_iB_i^{\mu})= c_i^{\mu}\neq 0$, the new interaction
  Hamiltonian between $s_{i}$ and $b_{i}$ can be defined as
  $V_{ib_i}=\sum_{\mu}A^\mu_i\otimes (B^\mu_{i}-c_{i}^{\mu})$. Then, one has
  $\tr(\tau_iB_i^{\prime\mu})=0$, where $B_i^{\prime\mu}=B_i^{\mu}-c_i^{\mu}$.
  The shifted Hamiltonian of the subsystem $s_i$ can be expressed as
  $H_i^{\prime}=H_i+\sum_{\mu}c_{i}^{\mu}A_i^\mu$~\cite{open4,krys}.
\end{thebibliography}
\end{document}